\documentclass{PoS}

\usepackage{subfigure}
\usepackage{verbatim}

\title{Strong Phase Measurements - Towards $\gamma$ at CLEO-c}

\ShortTitle{Strong Phase Measurements - Towards $\gamma$ at CLEO-c}

\author{\speaker{Andrew Powell} for the CLEO Collaboration \\
        University of Oxford, Denys Wilkinson Building, Oxford, OX1 3RH, United Kingdom\\
        E-mail: \email{a.powell1@physics.ox.ac.uk}}

\abstract{Strategies that utilise the interference effects within $B \to DK$ decays hold great potential for improving our 
sensitivity to the CKM angle $\gamma$. However, in order to exploit fully this potential, knowledge of parameters associated 
with the $D$ decay, such as strong-phase differences, are required. This essential information can be obtained from the unique 
quantum-correlated $\psi(3770)$ datasets at CLEO-c. Results of such analyses involving the decay modes $D \to K\pi, K\pi\pi^{0}, 
K\pi\pi\pi$ and $K^{0}_{S}\pi\pi$ will be presented.}

\FullConference{12th International Conference on B-Physics at Hadron Machines - BEAUTY 2009\\
		 September 07 - 11 2009\\
		 Heidelberg, Germany}

\begin{document}

\section{Introduction}
A theoretically clean method to extract the CKM-angle $\gamma$ is to exploit the interference present in $B^{\pm} \to DK^{\pm}$, 
where the $D$ is a $D^{0}$ or $\bar{D}^{0}$ decaying to a common final state, $f$. Decay rates in these channels are sensitive 
to the following amplitude ratios
\begin{eqnarray}
\frac{
A( B^{-}\!\to \bar{D}^{0} K^{-})}
{A( B^{-}\!\to D^{0} K^{-})} = r_{B}e^{i(\delta_{B} \, - \, \gamma)}, & {} &
\frac{
A( B^{+}\!\to D^{0} K^{+})}
{A( B^{+}\!\to \bar{D}^{0} K^{+})} = r_{B}e^{i(\delta_{B} \, + \, \gamma)}.
\label{eq:B-Params}
\end{eqnarray}
which are functions of three parameters: the ratio of the absolute magnitudes of the amplitudes, $r_{B}$; a $CP$-invariant 
strong-phase difference, $\delta_{B}$; and the weak phase $\gamma$. A variety of $\gamma$ extraction strategies have been 
suggested depending on the $D$ final state considered. For example, established final states include: two-body modes such as 
$K^{+}K^{-}$/$\pi^{+}\pi^{-}$ \cite{GL, GW}, $K^{\pm}\pi^{\mp}$~\cite{ADS}, as well as multi-body final states such as 
$K_{S}^{0}\pi^{+}\pi^{-}$~\cite{GGSZ, BP} and $K^{\pm}\pi^{\mp}\pi^{0}$/$K^{\pm}\pi^{\mp}\pi^{+}\pi^{-}$~\cite{AS}.\footnote{For 
a review of all these methods, and a summary of current and future $B^{\pm} \to DK^{\pm}$ $\gamma$ measurements, see 
Refs.~\cite{Miyabayashi} and \cite{Ricciardi}.} In all cases, the measurement of $\gamma$ is affected by properties of the $D$ 
decay amplitude. In order to exploit fully the sensitivity to the $B$-specific parameters ($r_{B}$, $\delta_{B}$ and $\gamma$) 
it is, therefore, highly advantageous to have prior knowledge of the parameters associated with the $D$ decay. This is where 
CLEO-c plays a crucial role.

These proceedings describe three sets of measurements performed by CLEO-c of $D$-specific parameters relevant to the measurement 
of $\gamma$. Sec.~\ref{sec:ADSParams} introduces the $D$ parameters of interest in the context of the $B$ decay rates. 
Sec.~\ref{sec:QuantumCorr} then explains how one can exploit quantum-correlations at the $\psi(3770)$ in order to probe these 
$D$ parameters. Sec.~\ref{sec:CLEO} describes the CLEO-c experiment and data sets used for the analyses. Secs.~\ref{sec:TQCA}, 
\ref{sec:Coherence} and \ref{sec:BinnedKspipi} describe the experimental procedure and results.  

\section{$D$ Parameters Associated with the ADS Method}
\label{sec:ADSParams}
In the case of the so-called \emph{ADS method}~\cite{ADS}, where $f = K^{\pm}\pi^{\mp}$, $D$-specific parameters contribute to 
the suppressed $B^{\pm}$ decay-rates as follows:
\begin{equation}
\Gamma( B^{\pm} \to ~(K^{\mp}\pi^{\pm})_{D}K^{\pm}) \propto  r_{B}^{2} + (r^{K\pi}_{D})^{2} + 2 \, r_{B} \, r^{K\pi}_{D} 
\cos(\delta_{B} + \delta^{K\pi}_{D} \pm \gamma),
\end{equation}
where $r^{K\pi}_{D}$ and $\delta^{K\pi}_{D}$ are analogous to the $B^{\pm}$ parameters $r_{B}$ and $\delta_{B}$; $r^{K\pi}_{D}$ 
is the absolute ratio of the doubly Cabibbo suppressed (DCS) to Cabibbo favoured (CF) amplitudes and $\delta^{K\pi}_{D}$ is the 
corresponding $D$ strong-phase difference. Futhermore, the extended method~\cite{AS}, which considers multi-body ADS modes i.e. 
$f = \{K^{\pm}\pi^{\mp}\pi^{0}$, \, $K^{\pm}\pi^{\mp}\pi^{+}\pi^{-} \}$, introduces an additional $D$ parameter: 
\begin{equation}
\Gamma( B^{\pm} \to ~(\bar{f})_{D}K^{-}) \propto  r_{B}^{2} + (r^{f}_{D})^{2} + 2 \, r_{B} \, r^{f}_{D} \, R_{f} \, 
\cos(\delta_{B} + \delta^{f}_{D} \pm \gamma),
\end{equation}
where $R_{f}$ is the coherence factor, and satisfies the condition $\{ R_{f} \in \mathbb{R}~\vert~ 0 \leq R_{f} \leq 1 \}$. This 
dilution term results from accounting for the resonant sub-structure of the multi-body mode. For modes whose intermediate 
resonances interfere constructively, $R_{f}$ tends to unity, however if the resonances interfere destructively, then $R_{f}$ 
tends to zero.

\section{Quantum Correlations at the $\psi(3770)$}
\label{sec:QuantumCorr}
Determination of strong-phase differences and coherence factors can be made from analysis of quantum-correlated 
$D^{0}\bar{D}^{0}$ pairs. Such an entangled state, with $C = -1$, is produced in $e^{+}e^{-}$ collisions at the $\psi(3770)$ 
resonance. To conserve this charge-conjugation state, the final state of the $D^{0}\bar{D}^{0}$ pair must obey certain selection 
rules. For example, both $D^{0}$ and $\bar{D}^{0}$ cannot decay to $CP$-eigenstates with the same eigenvalue. However, decays to 
$CP$-eigenstates of opposite eigenvalue are enhanced by a factor of two. More generally, final states that are accessible by 
both $D^{0}$ and $\bar{D}^{0}$ (such as $K^{-}\pi^{+}$) are subject to similar interference effects. Consequently, by 
considering time-integrated decay rates of double tag (DT) events, where both the $D^{0}$ and the $\bar{D}^{0}$ are 
reconstructed, one is sensitive to interference dependent parameters such as strong-phases and coherence factors. Furthermore, 
these decay rates are also sensitive to charm mixing. Charm mixing is described by two dimensionless parameters: $x \equiv 
(M_{1} - M_{2})/\Gamma$ and $y \equiv (\Gamma_{1} - \Gamma_{2})/2\Gamma$, where $M_{1,2}$ and $\Gamma_{1,2}$ are the masses and 
widths, respectively, of the neutral $D$ meson $CP$-eigenstates. The explicit dependence on the mixing parameters can be seen by 
considering the generalised, time-integrated, DT rate. That is, for a $D^{0}\bar{D}^{0}$ pair decaying to the final state 
$(f,\,g)$:
\begin{eqnarray}
\Gamma(f|g) & = & Q_{M}|A_{f}\bar{A}_{g} - \bar{A}_{f}A_{g}|^{2} + R_{M}|A_{f}A_{g} - \bar{A}_{f}\bar{A}_{g}|^{2} \, ,
\label{eq:GenDoubleTag}
\end{eqnarray}
where $A_{i}\equiv \left \langle i | D^{0} \right \rangle$, $\bar{A}_{i}\equiv \left \langle i | \bar{D}^{0} \right \rangle$. 
The coefficients $Q_{M}$ and $R_{M}$ posses the dependence on the mixing parameters, where $Q_{M} \equiv 1 -(x^{2} - y^{2})/2$ 
and $R_{M}\equiv (x^{2} + y^{2})/2$. 

\subsection{Probing strong-phases and coherence factors}
Letting $f$ represent the signal $D$ decay of interest, it is possible to obtain access to strong-phases and coherence factors 
by considering specific states of the `tag', $g$. As an example, we demonstrate here how sensitivity to strong-phases can be 
obtained by considering $g$ to be in a $CP$-eigenstate with eigenvalue $\lambda_{CP}$. For the purpose of this discussion, we 
simplify the problem by ignoring $D$-mixing effects, i.e. $x, y \to 0$. In this scenario, $Q_{M} \to 1$, $R_{M} \to 0$. 
Consequently, for $f=K^{-}\pi^{+}$, Eqn.(\ref{eq:GenDoubleTag}) reduces to:
\begin{eqnarray}
\Gamma(K^{-}\pi^{+}|CP) & \propto & | A_{K\pi}A_{CP} - \bar{A}_{K\pi}A_{CP}|^{2} \nonumber \\
{} & = &  |A_{K\pi}|^{2}|A_{CP}|^{2}\big(1 + (r^{K\pi}_{D})^{2} -2 \, \lambda_{CP} \, r^{K\pi}_{D} \cos(\delta^{K\pi}_{D}) 
\big).
\label{eq:KpiVsCP}
\end{eqnarray}
Therefore, with a knowledge of $|A_{K\pi}|$, $|A_{CP}|$ and $r^{K\pi}_{D}$, the observed asymmetry between the rates for 
$\lambda_{CP} = +1$ and $\lambda_{CP} = -1$ provides direct sensitivity to $\cos(\delta^{K\pi}_{D})$. When a multi-body signal 
mode is considered, such as $f = \{K^{\pm}\pi^{\mp}\pi^{0}$, \, $K^{\pm}\pi^{\mp}\pi^{+}\pi^{-} \}$, the amplitude $A_{f}$ must 
be integrated over all phase-space. This has the effect of modifying Eqn.~(\ref{eq:KpiVsCP}) through the transformation 
$\cos(\delta^{f}_{D}) \to R_{f}\cos(\delta^{f}_{D})$. Therefore, for $f = K^{-}\pi^{+}\pi^{0}$:
\begin{equation}
\Gamma(K^{-}\pi^{+}\pi^{0}|CP) = |A_{K\pi\pi^{0}}|^{2}|A_{CP}|^{2}\big(1 + (r^{K\pi\pi^{0}}_{D})^{2} -2 \, \lambda_{CP} \, 
r^{K\pi\pi^{0}}_{D} \, R_{K\pi\pi^{0}}\cos(\delta^{K\pi\pi^{0}}_{D}) \big).
\label{eq:KpipiVsCP}
\end{equation}
To give a more concrete overview, expressions from evaluating Eqn.~(\ref{eq:GenDoubleTag}) are listed in Table~\ref{tab:rates} 
for various tag modes against $f = K^{-}\pi^{+}$. As is demonstrated in Ref.\cite{Xing}, while $|A_{K\pi}|^{2}$ has direct 
correspondence to the CF branching fraction $(\mathcal{B}^{CF}_{K\pi})$,  $|\bar{A}_{K\pi}|^{2}$ and $|A_{CP}|^{2}$ possess 
dependence on the mixing parameters $x$ and $y$, i.e. $|\bar{A}_{K\pi}|^{2} = \mathcal{B}^{DCS}_{K\pi}(1 + \mathcal{O}(x,y))$. 
Consequently, a linear dependence on $x$ and $y$ is observed in some of the quantum correlated branching fractions quoted in 
Table~\ref{tab:rates}.

\begin{table}[h]
\centering
\begin{tabular}{l c}
\hline
\hline
{\bf Mode} & {\bf Relative Correlated Branching Fraction}\\
\hline
$K^{-}\pi^{+}$ vs. $K^{-}\pi^{+}$ & $R_{M}$\\
$K^{-}\pi^{+}$ vs. $K^{+}\pi^{-}$ & $(1+R_{W})^{2}-4r\cos\delta^{K\pi}_{D}(r\cos\delta^{K\pi}_{D} + y)$\\
$K^{-}\pi^{+}$ vs. $CP\pm$ & $1 + R_{WS} \pm 2r\cos\delta^{K\pi}_{D} + y$\\
$K^{-}\pi^{+}$ vs. $e^{-}$ & $1 - ry\cos\delta^{K\pi}_{D} -rx\sin\delta^{K\pi}_{D}$\\
$CP\pm$ vs. $CP\pm$ & 0\\
$CP+$ vs. $CP-$ & 4\\
$CP\pm$ vs. $e^{-}$ & $1 \pm y$\\
\hline
\hline
\end{tabular}
\caption{Correlated ($C = -1$) effective $D^{0}\bar{D}^{0}$ branching fractions to leading order in $x$, $y$ and $r^{2}$. The 
rates are normalised to the multiple of the uncorrelated branching fractions. Some rates show dependence to the wrong-sign rate 
ratio, $R_{WS} = r^{2} +ry' + R_{M}$, where $y' = (y\cos\delta^{K\pi}_{D} - x\sin\delta^{K\pi}_{D})$.}
\label{tab:rates}
\end{table}

\section{CLEO-c}
\label{sec:CLEO}
All measurements presented are made with $e^{+}e^{-} \to \psi(3770)$ data accumulated at the Cornell Electron Storage Ring 
(CESR). The CLEO-c detector was used to collect these data. Details of the experiment can be found elsewhere~\cite{CLEO}. The 
total integrated luminosity of the data is $818~{\rm pb}^{-1}$, however, only $281~{\rm pb}^{-1}$ have been used so far for the 
measurement of $\delta^{K\pi}_{D}$ presented in Sec.~\ref{sec:TQCA}.

\section{Measurement of the strong-phase difference in $D \to K^{-}\pi^{+}$}
\label{sec:TQCA}
The first analysis presented is that of the strong-phase difference in $D \to K^{-}\pi^{+}$. Implementing the method described 
in Ref.~\cite{AsnerSun}, this analysis has performed the first measurements of $y$ and $\cos(\delta^{K\pi}_{D})$ in 
quantum-correlated $\psi(3770)$ data. By comparing the correlated event yields, whose rates are listed in Table~1, with the 
uncorrelated expectations, we are able to extract $r^{2}$, $r\cos(\delta^{K\pi}_{D})$, $y$ and $x^{2}$. To achieve this, a 
knowledge of the relevant uncorrelated branching-ratios are needed. This information is gathered by averaging results of 
single-tagged yields at the $\psi(3770)$ with external measurements using incoherently-produced $D^{0}$ mesons. In addition, to 
extract $\cos(\delta^{K\pi}_{D})$ from $r\cos(\delta^{K\pi}_{D})$, knowledge of $r$ is required. This necessary information is 
obtained by including $R_{WS}$ and $R_{M}$ as external inputs to the least-squares fit. Furthermore, external measurements of 
the mixing parameters are used as constraints. All correlations amongst the inputs are accounted for.

The analysis has considered a total of seven $CP$-eigenstates reconstructed against the $K^{\pm}\pi^{\mp}$ signal mode: 
$K^{+}K^{-}$, $\pi^{+}\pi^{-}$, $K_{s}^{0}\pi^{0}$, $K_{s}^{0}\omega$, $K_{s}^{0}\pi^{0}\pi^{0}$, $K_{s}^{0}\eta$ and 
$K_{L}^{0}\pi^{0}$. In those DTs without a $K^{0}_{L}$, the signal is identified using two kinematic variables: the 
beam-constrained mass, $M \equiv \sqrt{E^{2}_{\rm Beam} - {\rm \bf p}^{2}_{D}}$, and $\Delta E \equiv E_{D} - E_{\rm Beam}$, 
where $E_{\rm Beam}$ is the beam energy, ${\rm \bf p}_{D}$ and $E_{D}$ are the $D^{0}$ candidate momentum and energy, 
respectively. The reconstruction of $K^{0}_{L}\pi^{0}$ events utilises the missing-mass technique described in Ref.~\cite{Qing}. 
The results of the fit are given in Table~2. The analysis finds a result of $\delta^{K\pi}_{D} = (22^{+11+9}_{-12-11})^{\circ}$, 
which is the first direct determination of this phase~\cite{TQCA}. An updated result following analysis of the full $818~{\rm 
pb}^{-1}$ dataset is in preparation.

\section{Measurement of the coherence factor and average strong-phase difference in $D \to K^{\pm}\pi^{\mp}\pi^{0}$ and $D \to 
K^{\pm}\pi^{\mp}\pi^{+}\pi^{-}$}
\label{sec:Coherence}
Determination of the average strong-phase difference and associated coherence factors for the modes $f = \{K\pi\pi^{0}, K3\pi 
\}$ have been made using an analogous technique to that described in Sec.~\ref{sec:TQCA}~\cite{Coherence}. As shown in 
Eqn.(\ref{eq:KpipiVsCP}), $CP$-tagged multi-body rates provide sensitivity to the product $R_{f}\cos(\delta^{f}_{D})$. A means 
of decoupling these parameters fortunately comes from considering the rate $\Gamma(f|f)$. Evaluating Eqn.(\ref{eq:GenDoubleTag}) 
for $g = f$, one obtains:
\begin{equation}
\Gamma(f|f) = Q_{M} |A_{f}|^{2} |\bar{A}_{f}|^{2} \, \Big(1 - (R_{f})^{2} \Big) + |A_{f}|^{4} \, R_{M} \,\Big( 1 
-2(r^{f}_{D})^{2} + (r^{f}_{D})^{4} \Big).
\label{eq:LSRate}
\end{equation}
In the case of the two-body mode, $f=K^{\pm}\pi^{\mp}$, $R_{f}=1$ and Eqn.(\ref{eq:LSRate}) reduces to $|A_{f}|^{4}R_{M}$ as 
quoted in Table~1. However, for multi-body final states, one observes that $(1 - R_{f}^{2})$ is the leading term in 
Eqn.(\ref{eq:LSRate}). Consequently, the rate $\Gamma(f|f)$ provides direct sensitivity to $R_{f}$ and allows for a decoupling 
of the parameters. All the $CP$-tags listed in Sec.~\ref{sec:TQCA} are employed in this analysis, as well as $K^{0}_{S}\phi$, 
$K^{0}_{S}\eta'$ and $K^{0}_{L}\omega$.

\begin{figure}[h]
\begin{minipage}[h]{.49 \textwidth}
\includegraphics[width = \textwidth]{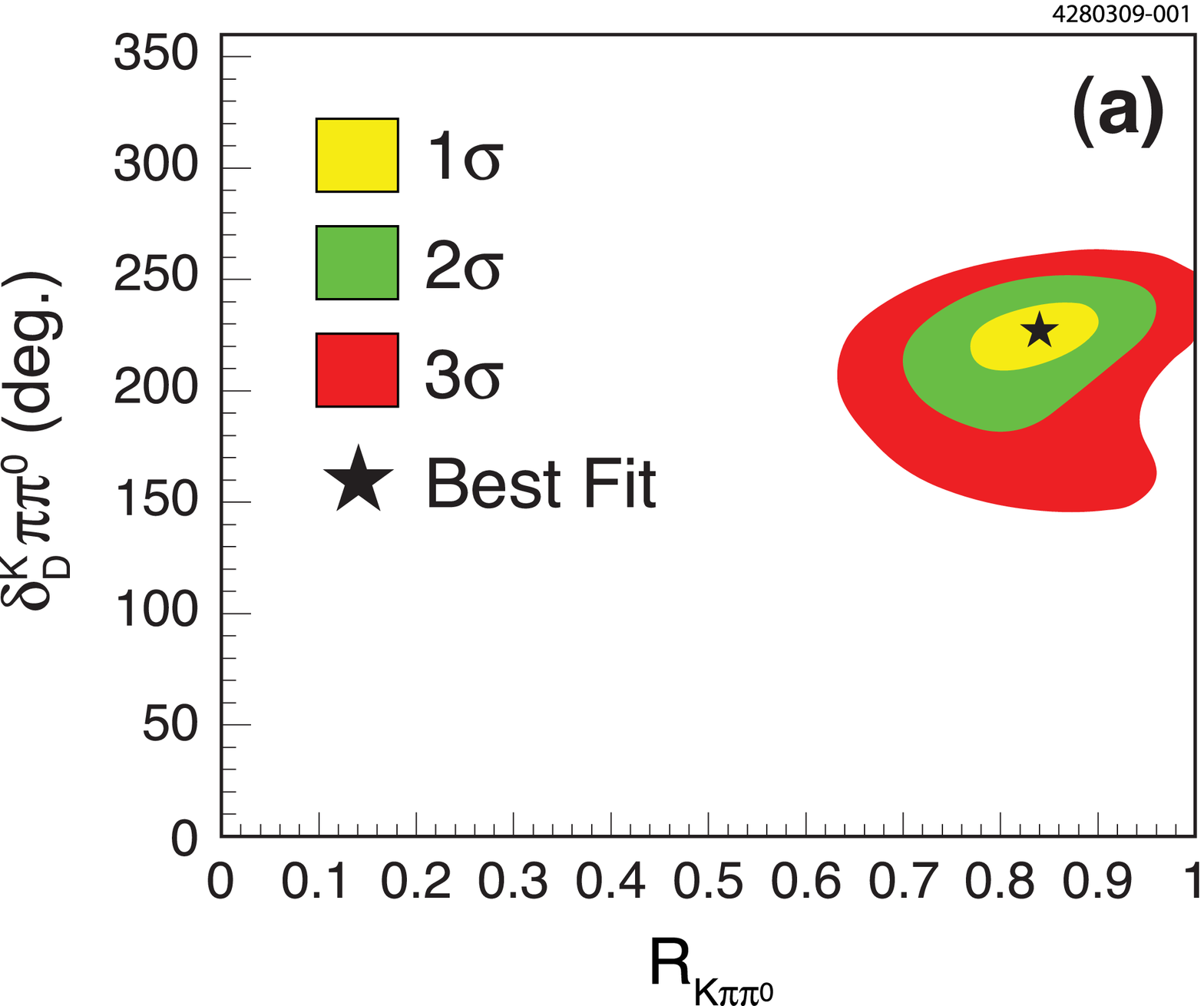}
\end{minipage}
\hspace{2mm}
\begin{minipage}[h]{.49 \textwidth}
\includegraphics[width =
\textwidth]{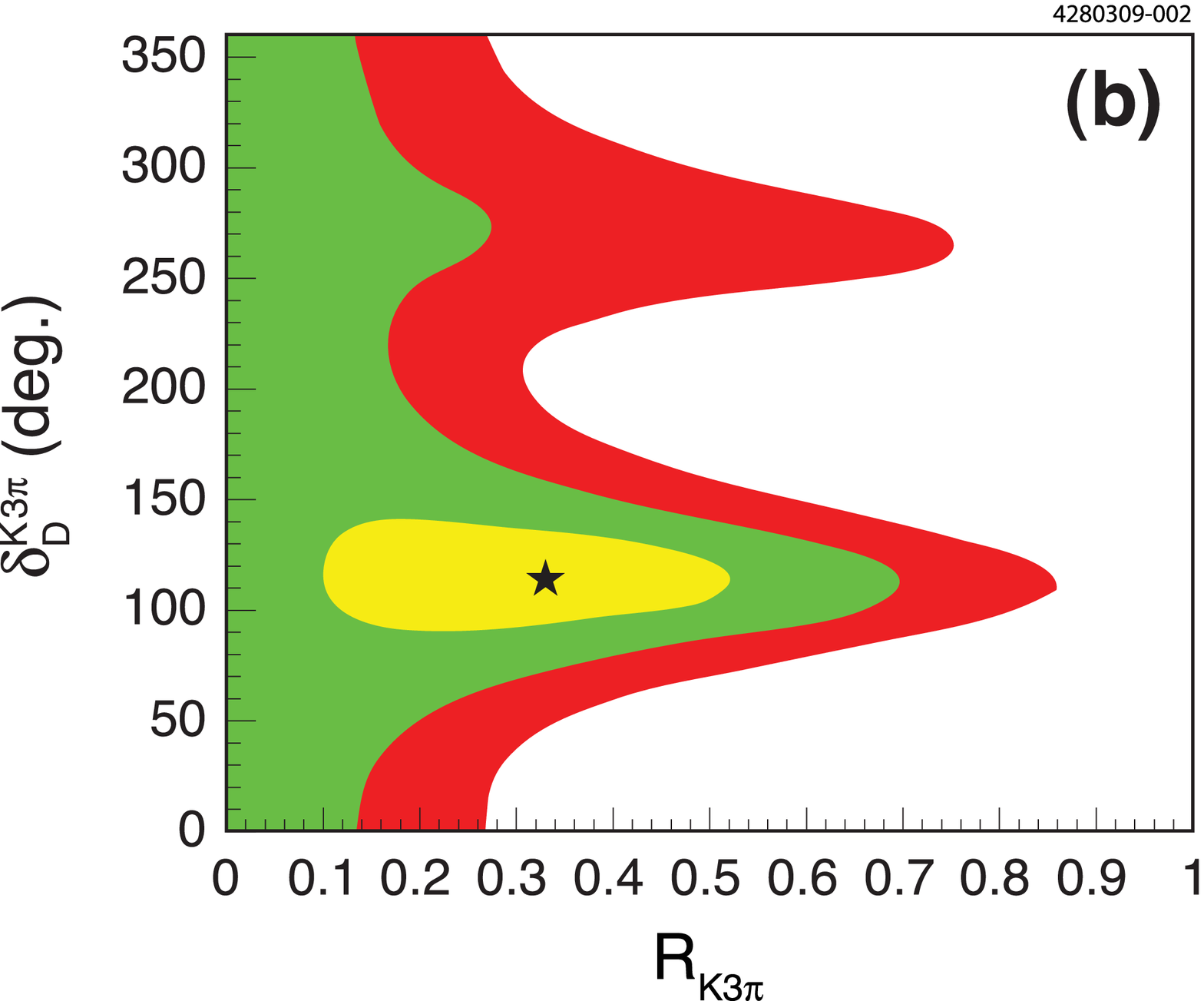}
\end{minipage}
\caption{The limits determined on (a) ($R_{K\pi\pi^{0}},\,  \delta^{K\pi\pi^{0}}_{D}$) and (b) ($R_{K3\pi},\,  
\delta^{K3\pi}_{D}$) at the 1, 2 and $3\sigma$ levels.}
\label{fig:coherenceContours}
\end{figure}

As was done in the $K^{\pm}\pi^{\mp}$ analysis, a least-squares fit has been used to extract both mixing and strong-phase 
parameters. Likelihood contours in $R_{f}$, $\delta^{f}_{D}$ parameter space are shown in Fig.~\ref{fig:coherenceContours}(a) 
for $f = K\pi\pi^{0}$, and Fig.~\ref{fig:coherenceContours}(b) for $f = K3\pi$. The best-fit values of the coherence factors and 
average strong-phases are $R_{K\pi\pi^{0}} = 0.84 \pm 0.07$, $\delta^{K\pi\pi^{0}}_{D} = (227^{+14}_{-17})^{\circ}$, $R_{K3\pi} 
= 0.33^{+0.20}_{-0.23}$ and $\delta^{K3\pi}_{D} = (114^{+28}_{-23})^{\circ}$. The uncertainties quoted are a combination of 
statistical and systematic errors. 

The results show significant coherence for $D^{0} \to K\pi\pi^{0}$, but much less so for $D^{0} \to K\pi\pi\pi$. These results 
will improve the measurement of $\gamma$ and the amplitude ratio $r_{B}$ in $B^{\pm} \to DK^{\pm}$, where the $D$ decays to 
$K\pi\pi^{0}$ and $K\pi\pi\pi$. Earlier preliminary results of $R_{K3\pi}$ and $\delta^{K3\pi}_{D}$~\cite{LakeLouise} combined 
with CLEO-c's measurement of $\delta^{K\pi}_{D}$ were shown to improve the expected sensitivity to $\gamma$ at LHCb in a 
combined ADS analysis of $K\pi$ and $K\pi\pi\pi$ final states by up to $40\%$~\cite{LHCbADS}.

\section{Measurement of strong-phase variations in $D \to K^{0}_{S}\pi^{+}\pi^{-}$}
\label{sec:BinnedKspipi}
The current best constraints on $\gamma$ come from measurements in $B^{\pm} \to D(K^{0}_{S}\pi^{+}\pi^{-})K^{\pm}$ and related 
modes~\cite{Belle, Babar} by performing likelihood fits to the $K^{0}_{S}\pi^{+}\pi^{-}$ Dalitz plot~\cite{GGSZ}. These fits 
require models to represent the $D^{0} \to K^{0}_{S}\pi^{+}\pi^{-}$ resonant amplitude structure. Since these models possess 
certain assumptions, an inherent systematic uncertainty is associated with this technique. Current estimates predict this error 
to be between $5^{\circ}$ and $9^{\circ}$, meaning the $\gamma$ measurement would soon become systematically limited at the next 
generation of flavour-physics experiment. However, an alternative \emph{model-independent} method has been proposed where events 
are counted in specified regions of the $K^{0}_{S}\pi^{+}\pi^{-}$ Dalitz plot~\cite{GGSZ,BP}, thus eliminating the 
model-uncertainty. This method relies on necessary strong-phase parameters having been determined at CLEO-c.

As Dalitz plot variables we use the invariant-mass squared of the $K^{0}_{S}\pi^{-}$ and $K^{0}_{S}\pi^{+}$ pairs, which we 
label as $s_{-}$ and $s_{+}$, respectively. The strong-phase at a given point in the $K^{0}_{S}\pi^{+}\pi^{-}$ Dalitz plot is 
then $\delta_{D}(s_{-}, s_{+})$. For the phase difference between $D^{0} \to K^{0}_{S}\pi^{+}\pi^{-}$ and $\bar{D}^{0} \to 
K^{0}_{S}\pi^{+}\pi^{-}$ at the same point in the Dalitz plot, we define
\begin{equation}
\Delta\delta_{D} \equiv \delta_{D}(s_{-}, s_{+}) - \delta_{D}(s_{+}, s_{-}).
\end{equation}
The quantities measured by CLEO-c that provide input to the model-independent $\gamma$ determination are the averages of 
$\cos(\Delta\delta_{D})$ and $\sin(\Delta\delta_{D})$ in the $i$th Dalitz plot bin. We denote these terms $c_{i}$ and $s_{i}$, 
respectively. In a completely analogous manner to the analyses presented in Secs.~\ref{sec:TQCA} and \ref{sec:Coherence}, 
$c_{i}$ can be determined from $CP$-tagged decay rates, while $s_{i}$ is extracted from considering the double Dalitz plot of 
$K^{0}_{S}\pi^{+}\pi^{-}$ vs. $K^{0}_{S}\pi^{+}\pi^{-}$. Furthermore, additional constraints on $c_{i}$ and $s_{i}$ are obtained 
through $K^{0}_{L}\pi^{+}\pi^{-}$ events. 

\begin{figure}[h]
\begin{center}
\subfigure[]{\label{fig:Kspipi-a}
\includegraphics[width = 0.47 \textwidth]{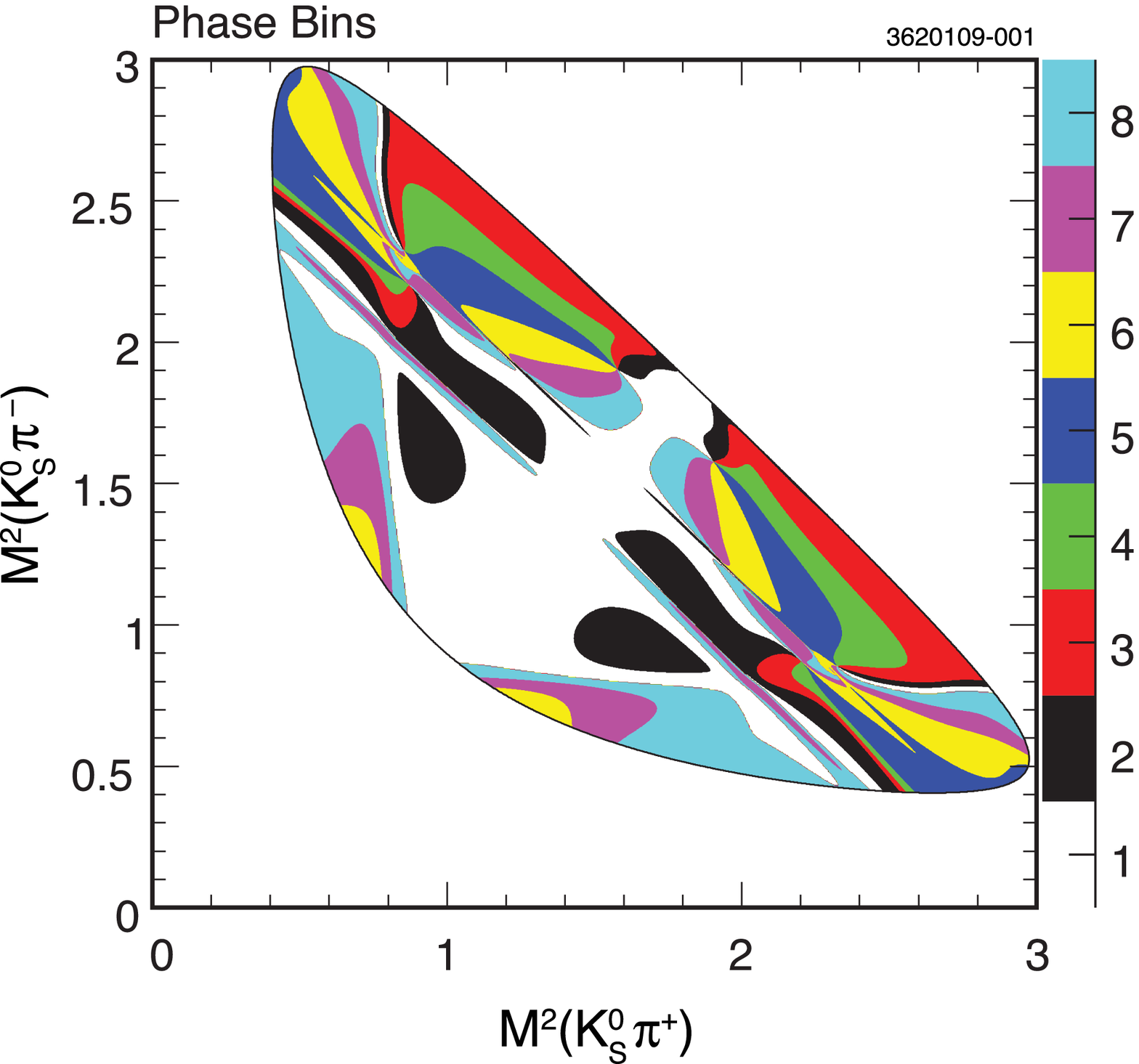}}
\subfigure[]{\label{fig:Kspipi-b}
\includegraphics[width = 0.47 \textwidth]{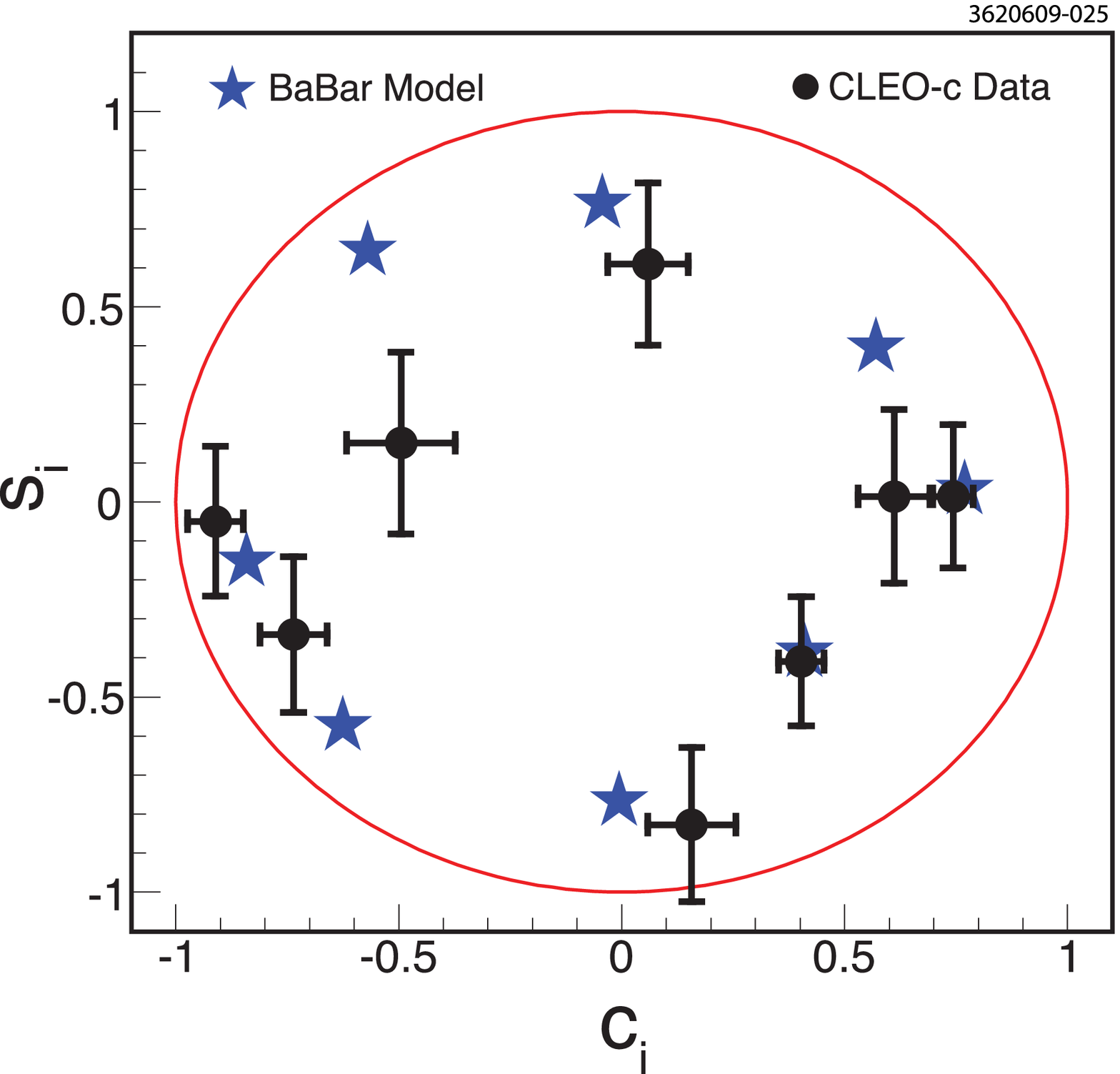}}
\end{center}
\vspace{-5mm}
\caption{In (a), the uniform $|\Delta\delta_{D}|$ binning of the $K^{0}_{S}\pi^{+}\pi^{-}$ Dalitz plot. In (b), the comparison 
of the measured $c_{i}$ and $s_{i}$ (circles with error bars) to the predictions from the BaBar isobar model (stars).}
\label{fig:Kspipi}
\end{figure}

The choice of Dalitz plot binning affects the statistical precision of the analysis. It has been demonstrated in Ref.~\cite{BP} 
that it is beneficial to choose bins such that $\Delta\delta_{D}$ varies as little as possible across each bin. The binning used 
in this analysis, with eight-pairs of bins uniformly dividing $\Delta\delta_{D}$ over the range $[0, 2\pi]$, is shown in 
Fig.~\ref{fig:Kspipi-a}. The location of these bins in phase space is evaluated from referring to the BaBar isobar model given 
in Ref.~\cite{BaBarIsobar}.

The values of $c_{i}$ and $s_{i}$ from the combined analysis of $K^{0}_{S}\pi^{+}\pi^{-}$ and $K^{0}_{L}\pi^{+}\pi^{-}$ tagged 
events are shown graphically in Fig~\ref{fig:Kspipi-b}. When used as input to the $\gamma$ measurement, these results are 
expected to replace the current model uncertainty of $5^{\circ} - 9^{\circ}$ with an uncertainty due to the statistically 
dominated error on $c_{i}$ and $s_{i}$ of $1.7^{\circ}$~\cite{QingKspipi}. 

\section{Conclusion}
The importance of CLEO-c's quantum-correlated $\psi(3770)$ dataset in the context of measuring the CKM angle $\gamma$ has been 
described. Analysis of a variety of two- and multi-body $D^{0}$ decays with these data have provided vital measurements of 
$D^{0}$ strong-phases, and associated parameters, for model-independent $\gamma$ measurements at LHCb. In addition to the modes 
presented here, results are in preparation for other promising final states, such as $D^{0} \to K^{0}_{S}K^{+}K^{-}$.

\begin{thebibliography}{99}

\bibitem{GL} M. Gronau and D. London, Phys. Lett. B {\bf 253} (1991) 483.
\bibitem{GW} M. Gronau and D. Wyler, Phys. Lett. B {\bf 265} (1991) 172.
\bibitem{ADS} D. Atwood, I. Dunietz and A. Soni, Phys. Rev. Lett. {\bf 78} (1997) 3257.
\bibitem{GGSZ} A. Giri, Y. Grossman, A. Soffer and J. Zupan, Phys. Rev. D {\bf 68} (2003) 054018.
\bibitem{BP} A. Bondar and A. Poluektov, Eur. Phys. J. {\bf 47} (2006) 347.
\bibitem{AS} D. Atwood and A. Soni, Phys. Rev. D {\bf 68} (2003) 033003.
\bibitem{Miyabayashi} K. Miyabayashi, these proceedings.
\bibitem{Ricciardi} S. Ricciardi, these proceedings.
\bibitem{Xing} Z.Z. Xing, Phys. Rev. D {\bf 55} (1997) 196.
\bibitem{CLEO} Y.~Kubota {\it et al.}, Nucl. Instrum. Meth. Phys. Res., Sect. A {\bf 320}, (1992) 66; D.~Peterson {\it et al.}, 
Nucl. Instrum. Meth. Phys. Res., Sect. A {\bf 478}, (2002) 142.
\bibitem{AsnerSun} D. Asner and W. Sun, Phys. Rev. D {\bf 73}, (2006) 034024; {\bf 77} (2008) 019901(E).
\bibitem{Qing} Q. He \emph{et al.}[CLEO Collaboration], Phys. Rev. Lett. {\bf 100} (2008) 091801.
\bibitem{TQCA} J. Rosner \emph{et al.} [CLEO Collaboration], Phys. Rev. Lett. {\bf 100} (2008) 221801, D. Asner \emph{et al.} 
[CLEO Collaboration] Phys. Rev.D {\bf 78} (2008) 012001.
\bibitem{Coherence} N. Lowery \emph{et al.} [CLEO Collaboration], Phys. Rev. D {\bf 80} (2009) 031105(R).
\bibitem{LakeLouise} A. Powell [CLEO Collaboration], in \emph{Fundamental Interactions: Proceedings of the 23rd Lake Louise 
Winter Institute 2008}, World Scientific Publishing.
\bibitem{LHCbADS} K. Akiba \emph{at al.} CERN Report No. CERN-LHCb-2008-031 (2008).
\bibitem{Belle} K .Abe \emph{et al.} [Belle Collaboration], arXiv:0803.3375[hep-ex]
\bibitem{Babar} B. Aubert \emph{et al.} [BaBar Collaboration], Phys. Rev. D {\bf 78} (2008) 034023.
\bibitem{BaBarIsobar} B. Aubert \emph{et al.} [BaBar Collaboration], arXiv:hep-ex/0607104.
\bibitem{QingKspipi} R. Briere \emph{et al.} [CLEO Collaboration], Phys. Rev. D {\bf 80} (2009) 032002.

\end{thebibliography}
\end{document}